\documentclass[12pt]{article}
\pdfoutput=1
\usepackage{graphics}
\usepackage{graphicx}
\usepackage{mathrsfs}
\usepackage{amssymb}
\usepackage{amsmath}
\usepackage{verbatim}
\usepackage{float}
\usepackage{slashed}
\usepackage{bbm}
\newcommand{\be}{\begin{equation}}
\newcommand{\ee}{\end{equation}}
\newcommand{\bea}{\begin{eqnarray}}
\newcommand{\eea}{\end{eqnarray}}

\def\4vol{{\int d^4x \sqrt{-g}}}

\def\beq{\begin{equation}}
\def\eeq{\end{equation}}
\def\bea{\begin{eqnarray}}
\def\eea{\end{eqnarray}}
\def\bitem{\begin{itemize}}
\def\eitem{\end{itemize}}
\def\ba{\begin{array}}
\def\ea{\end{array}}
\def\bal{\begin{align}}
\def\eal{\end{align}}
\def\bi{\begin{itemize}}
\def\ei{\end{itemize}}

\newcommand{\nc}{\newcommand}
\nc{\nt}{\tilde{N}}
\nc{\ra}{\rightarrow}
\nc{\lsim}{\begin{array}{c}\,\sim\vspace{-21pt}\\< \end{array}}
\nc{\gsim}{\begin{array}{c}\sim\vspace{-21pt}\\> \end{array}}
\nc{\tnt}{\tilde{N}}
\nc{\tst}{\tilde{t}}
\nc{\gev}{\,{\rm GeV}}
\nc{\mev}{\,{\rm MeV}}
\nc{\ev}{\,{\rm eV}}
\nc{\kev}{\,{\rm keV}}
\nc{\tev}{\,{\rm TeV}}

\nc{\mz}{M_Z}
\nc{\mpl}{M_{\rm pl}}
\nc{\mw}{m_{\rm weak}}
\nc{\mh}{m_{\rm h}}
\nc{\gw}{\hat{g}_2}
\nc{\gy}{\hat{g}_Y}
\nc{\gst}{\hat{g}_3}
\nc{\lambdah}{\hat{\lambda}}
\nc{\muh}{\hat{\mu}}
\nc{\vh}{\hat{v}}
\nc{\Bkk}{B^{(1)}}
\nc{\Wkk}{W^{(1)}}
\nc{\Wpmkk}{W_{\pm}^{(1)}}
\nc{\Zkk}{W^{3(1)}}
\nc{\gakk}{\gamma^{(1)}}
\nc{\Hkk}{H^{(1)}}
\nc{\hkk}{h^{(1)}}
\nc{\akk}{a_0^{(1)}}
\nc{\apmkk}{a_{\pm}^{(1)}}
\nc{\Bz}{B^{(0)}}
\nc{\Wz}{W^{(0)}}
\nc{\Wpmz}{W_{\pm}^{(0)}}
\nc{\Zz}{B^{(0)}}
\nc{\gaz}{\gamma^{(0)}}
\nc{\Hz}{H^{(0)}}
\nc{\hz}{h^{(0)}}
\nc{\mwh}{\hat{m}_W}
\nc{\mbh}{\hat{m}_B}

\nc{\ie}{{\it i.e.~}}          \nc{\etal}{{\it et al.~}}
\nc{\eg}{{\it e.g.~}}          \nc{\etc}{{\it etc.~}}
\nc{\cf}{{\it c.f.~}}

\nc{\LL}{L}
\nc{\vv}{\tilde{v}}

\title{
\vspace*{5mm} \Large\textbf{Hints of R-parity violation in $B$ decays into 
$\tau \nu$}
\vspace*{1.0cm}
\author{\textbf{N.~G.~Deshpande$^a$, A.~Menon$^a$}\\
\normalsize\emph{$^a$Institute of Theoretical Science, University of
Oregon, Eugene, OR 97403}\\
}
}                
\date{\today}
\begin{document}
\setcounter{page}{0}
\maketitle
\begin{abstract}
In this article we show that the recently observed enhanced semi-leptonic and 
leptonic decay rates of the $B$ meson into $\tau \nu$ modes can be explained 
within the frame work of R-parity violating (RPV) MSSM. In particular, RPV 
contributions involving the exchange of right-handed down-type squarks can
give a universal contribution to the $B^+ \to \tau \nu$, $B \to D \tau \nu$
and the $B \to D^* \tau \nu$ decays in the model we propose. We find that the masses and 
couplings that explain the enhanced $B$ decay rates are phenomelogically viable and 
the squarks can possibly be observed at the LHC.
\end{abstract}

\thispagestyle{empty}
\newpage
\setcounter{page}{1}

\section{Introduction}
\label{sec:Intro}

The Babar collaboration~\cite{Lees:2012xj} has reported recently improved 
measurements of the ratios of the semileptonic decays of $\frac{
\mathcal{BR}(B \to D \tau \nu)}{\mathcal{BR}(B \to D l \nu)}$
and $\frac{\mathcal{BR}(B \to D^* \tau \nu)}{\mathcal{BR}(B \to D^* l \nu)}$.
Both these observation exceed the Standard Model (SM) expectations by more
than $2\sigma$. Furthermore, the Babar collaboration~\cite{Lees:2012ju} also
finds the branching ratio $B \to \tau \nu$ exceeds the SM prediction 
obtained using the 
global fit of the CKM matrix. We summarize the experimental measurements: 
\bea
R^{\rm exp}(D) &=& 
\frac{\mathcal{BR}(B \to D \tau \nu)}{\mathcal{BR}(B \to D l \nu)} = 0.440 \pm 
0.072 \label{BDtnu} \\
R^{\rm exp}(D^*) &=& 
\frac{\mathcal{BR}(B \to D^* \tau \nu)}{\mathcal{BR}(B \to D^* l \nu)} = 0.332
\pm 0.030 \label{BDstartnu} \\
\mathcal{BR}(B \to \tau \nu) &=& (1.67 \pm 0.3) \times 10^{-4}. \label{Btnu}
\eea
where for $B \to \tau \nu$ we used the HFAG average value~\cite{Asner:2010qj}.
For comparison the expected SM values are~\cite{Kamenik:2008tj,
Fajfer:2012vx}:
\bea
R^{\rm SM}(D) &=& 0.297 \pm 0.017\\
R^{\rm SM}(D*) &=& 0.252 \pm 0.003 
\eea
and we take the more conservative prediction based on the average $|V_{ub}|$
given in Ref.~\cite{pdg} for 
\bea
\mathcal{BR}(B \to \tau \nu)^{\rm SM} &=& (1.04 \pm 0.31) \times 10^{-4}
\eea

We therefore find that each of these rates are enhanced over the Standard Model
expectation.
Clearly if these deviations from the SM are taken seriously there has to be
new physics involved which distinguishes the $\tau$-lepton from the lighter
lepton families, $\mu$ and $e$. We also know from decays of the $Z$ and $W$
gauge bosons that universality of leptons is obeyed to a high degree of 
precision in those decays. Thus exchange of new particles that preferentially 
couple to the 
third family are probably necessary.

Many theoretical attempts have been made recently to understand the above 
discrepencies. Purely phenomenological studies based on four 
fermion operators have been carried out in Ref.~\cite{Datta:2012qk,Fajfer:2012jt,Becirevic:2012jf}. Specific models
involving extra Higgs doublets~\cite{Crivellin:2012ye} 
and leptoquarks~\cite{Fajfer:2012jt} have been discussed.

In this note we consider a model involving R-parity violating interactions
in the superpotential previously studied in 
Ref.~\cite{Erler:1996ww,Grossman:1995gt,Aida:2010xi,Bose:2011tb}. In particular, down-type squark, 
$\tilde{d}$, exchange
provides a mechanism to explain all the data reasonably well.  
Our model has the further virtue of providing a universal 
explanation for all the enhancements in the processes involving the
transition $b \to (c,u) \tau \nu$.

\section{R-parity violating MSSM}
\label{sec:RPV}

In this section we will give a brief overview of the R-parity violating (RPV)
MSSM and discuss the possibility that it can explain the enhanced decay rates
of the $B$ meson. We will also briefly discuss the possibility of observing
this scenario at the LHC.

\subsection{Setup} \label{sec:setup}

Once we allow for R-parity violating operators, the down-type Higgs chiral 
superfield $H_d$ and the three lepton superfields $L_i$ cannot be 
distinguished. Hence the minimal R-parity violating superpotential terms
allowed by $SU(3)_c \times SU(2)_L \times U(1)_Y$ gauge invariance 
are~\cite{Barbier:2004ez,Chemtob:2004xr}
\bea
W_{\rm RPV} = \mu_i L_i H_u + \frac{1}{2} \lambda_{ijk} L_i L_j 
E_k + \lambda'_{ijk} L_i Q_j D_k + \frac{1}{2} \lambda''_{ijk} U_i D_j D_k
\eea
where the generation indices $(i,j,k) \in (1,2,3)$ for the 
$Q,U,D,E$ superfields and $(i,j)=(1,2,3,4)$ for $L$ superfield.
$SU(2)_L$ and $SU(3)_C$ gauge invariance implies $\lambda_{ijk} = - \lambda_{jik}
$ and $\lambda''_{ijk} =-\lambda''_{ikj}$. In addition to these superpotential 
RPV operators, R-parity violating soft-supersymmetry breaking operators are 
also present. As we will be working at low $\tan \beta$ these operators give 
sub-dominant contributions to the semi-leptonic decays of the $b$-quark and
are not of interest in this article.

The strong constraints from 
proton decay can be avoided by imposing the discrete $Z_3^B$ baryon 
number symmetry discussed in Ref.~\cite{Ibanez:1991pr}. The charge 
assignments for each of the MSSM chiral superfields are shown in 
Tab.~\ref{tab:Z3B}
\begin{table}
\begin{center}
\begin{tabular}{|c|c|c|c|c|c|c|}
\hline
Chiral Field & $Q$ & $U$ & $D$ & $L$ & $E$ & $H_u$ \\
\hline
$Z_3^B$ charge & 0 & -1 & 1 & -1 & -1 & 1 \\
\hline  
\end{tabular}
\caption{$Z_3^B$ charge assignments of the MSSM particle content}
\label{tab:Z3B}
\end{center}
\end{table}
where a chiral superfield $\Phi$ transforms as $\Phi \to \Phi e^{2\pi i \phi/N}$
for $Z_3^B$ charge $\phi$. The $Z_3^B$ symmetry explicitly excludes the 
$\lambda''$ proportional terms. Furthermore in order to avoid experimental
and cosmological constraints on neutrino mass we assume that both the bilinear
RPV term $\mu_i$ and the related soft-SUSY breaking term are small and hence
their phenomenological impact on b-decays are negligible. Therefore under
these simplifying assumptions and rotating in to the basis of the physical 
$H_d$ the total superpotential has the form
\bea
W &=& W_{\rm MSSM} + W_{\rm RPV} \\
W_{\rm RPV} &=& \frac{1}{2} \hat \lambda_{ijk} \hat L_i \hat L_j \hat E_k^c + 
\hat \lambda'_{ijk} \hat E_i \hat Q_j \hat D_k^c
\eea
Now rotating into the mass eigenbasis of the charged leptons and the quarks we 
find
\bea
W_{\rm RPV} &=& \frac{1}{2} \lambda_{ijk}(N_i E_j - E_i N_j) E_k^c + 
\lambda'_{ijk} (N_i D_j - V_{jl}^{\rm CKM} E_i U_l )D_k^c \label{WRPV}
\eea
where the un-hatted $\lambda$ and $\lambda'$ are the effective RPV couplings 
in this basis.\footnote{We have chosen to leave the neutrinos in the gauge basis
as the final state neutrino in the operator expansion is also in the gauge 
basis.} Eq.~(\ref{WRPV}) is quite general and involves a suitable redefinition
of $\lambda$ and $\lambda'$, where $\hat \lambda$ and $\hat \lambda'$ are 
related to $\lambda$ and $\lambda'$ by the rotation matrices. For an explicit
derivation see Eq.~(6.1) of Ref.~\cite{Barbier:2004ez}. This is the appropriate base in which
to discuss the flavor problem.

It is perfectly possible to carry out the B-decay phenomenology based on 
Eq.~(\ref{WRPV}). We are however proposing a more restrictive model in the 
following discussion. Our primary motivation to do this, is to provide a 
universal explanation for the enhancements seen in the experiments in 
Eqs.~(\ref{BDtnu}), (\ref{BDstartnu}) and (\ref{Btnu}). To motivate this model,
we shall assume a Froggatt-Nielsen like mechanism~\cite{rpv_model} for 
generating the flavor heirarchy. For example, we can assume an abelian flavor 
symmetry $U(1)_X$ under which the $H_u Q_3 U_3$ operator is invariant~\cite{Barbier:2004ez} 
providing a rationale for a heavy top. This $U(1)_X$ symmetry is broken by an MSSM singlet 
field acquiring a vacuum expectation value (VEV), $\theta$, thereby generating the down-type 
Yukawa terms ($\hat \lambda_{ij}^d H_d Q_i D_j$), where
\bea
\hat \lambda_{ij}^d = y^d_{ij} \left(\frac{\theta}{M} \right)^{h_d + q_i + d_j}
\eea
and the correlated RPV violating couplings are
\bea
\hat \lambda'_{ijk}  = y^d_{jk} \left(\frac{\theta}{M} \right)^{l_i + q_j + d_k} = \left(\frac{\theta}{M} \right)^{l_i - h_d} \hat \lambda_{jk}^d
\eea
where $M$ is messenger mass scale of flavor symmetry breaking and $(l,q,d,h_d)$
are the $U(1)_X$ charges of the $(L,Q,D,H_d)$ superfields respectively. Notice
that the flavor structure of the RPV violating operators is the same as 
the Yukawas, so that the Super-GIM mechanism will still suppress flavor 
violation.\footnote{If we assume that the RPV violating terms are proportional to 
$\hat \lambda^d$ at the scale
$M$, RG evolution to the low scale still preserves the flavor structure. However
this alignment may be destroyed by the evolution in the bilinear RPV terms. Hence for
simplicity we assume that this alignment is only true at the low energy scale where
by assumptions the bilinear terms are negligible.}  
Hence for $\lambda'_{333} \sim \mathcal{O}(1)$ we need either $H_d Q_3
 D_3$ to be invariant under $U(1)_X$ and the ratio of the Higgs VEVs, $\tan \beta 
= v_u/v_d$, is large or $l_3 \lsim h_d$ and $\tan \beta$ is small. 
In this simple picture for either of the 
above cases, when we rotate into the quark mass basis, \emph{the only non-zero terms} 
are $\lambda_{333}',\lambda_{322}',\lambda_{311}'$ and their ratios are
\bea
\frac{\lambda'_{322}}{\lambda'_{333}} = \frac{m_s}{m_b} \sim 0.02 \mbox{ and }
\frac{\lambda'_{311}}{\lambda'_{333}} = \frac{m_d}{m_b} \sim 2 \times 10^{-3}.
\label{eq:lprimeheirarchy}
\eea
The RPV couplings $\lambda$ are similarly related to the lepton Yukawas and therefore will be
suppressed compared to the $\lambda'$. Further, in order to avoid a landau pole in the 
the RPV coupling, $\lambda'$, we assume that $\lambda'_{333} \lsim 1.07$~\cite{Barger:1995qe}. We reiterate that the structure of these couplings is due to
our choice of the mechanism for the break down of the $U(1)_X$ flavor
symmetry of the RPV model.

\subsection{RPV contributions to $B$ rare decays}

\begin{figure}
\centering
\includegraphics[width=0.7\textwidth]{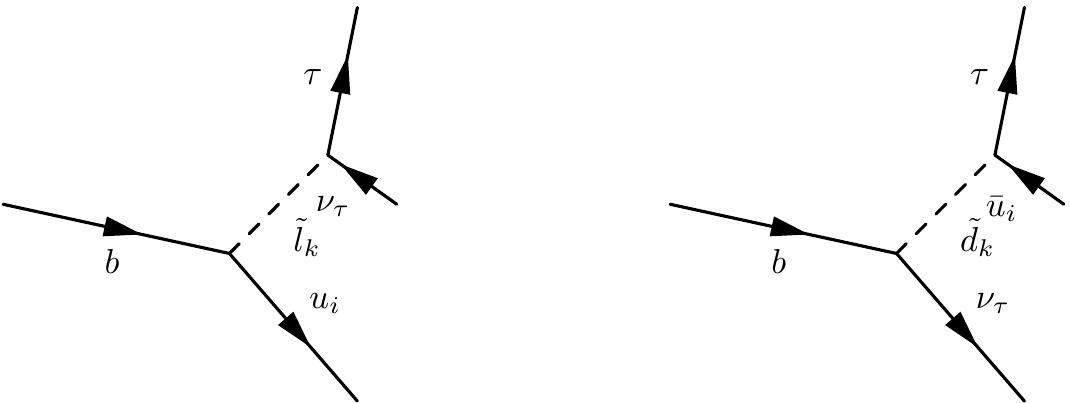}
\label{fig:feyndiag}
\caption{Squark and slepton induced rare b-quark decays}
\end{figure}

Using Eq.~(\ref{WRPV}) the RPV fermion-fermion-sfermion interaction terms are
\bea
\mathcal{L}_{RPV} &\subset& - \lambda'_{ijk} \left(\tilde{{\nu}}_{iL}
\bar{d}_{kR} d_{jL} + \tilde{d}_{jL} \bar{d}_{kR} {\nu}_{iL} + 
\tilde{d}_{kR}^* \bar{{\nu}}_{iR}^c d_{jL} -  V^{\rm CKM}_{jl} \left(\tilde{{l}}_{iL} 
\bar{d}_{kR} u_{lL} + \tilde{u}_{lL} \bar{d}_{kR} l_{iL} +
\tilde{d}_{kR}^* \bar{{l}}_{iR}^c u_{lL} \right) \right)
\nonumber \\
& &-\frac{1}{2} \lambda_{ijk} \left(\tilde \nu_{iL} 
\bar{l}_{kR} l_{jL} + \tilde l_{jL} \bar{l}_{kR} \nu_{iL} + \tilde l_{kR}^*
\bar{\nu}_{iR}^c l_{jL} - (i\leftrightarrow j)\right) + {\rm h.c.}
\eea
Hence the two possible contributions to the semi-leptonic decays of 
the $B$ meson are due to the exchange of sleptons and 
down-type squarks shown in Fig.~\ref{fig:feyndiag}.
Integrating out the sleptons and squarks we find the 4-fermi interaction terms
and keeping only the leading $\lambda$ and $\lambda'$ terms  
\bea
\mathcal{L}_{4\rm f} \subset-  V_{sm}^{\rm CKM} &&\left[
\lambda_{ijk} \lambda^{'*}_{rst} \left(\frac{V_{jl}^LV_{rl}^{*L}}{m_{\tilde l_l}^2}
\right) (\bar l_k P_L \nu_i) (\bar u_m P_L d_t) + \nonumber \right. \\
&& \lambda_{ijk} \lambda'_{rst}
\left(\frac{V_{(k+3)l}^{*L} V_{rl}^{L}}{m_{\tilde l_l}^2}
\right) (\bar \nu_i C P_L l_j) (\bar d_t P_L u_m) + \nonumber\\
&& \lambda'_{ijk} \lambda'_{rst}
\left(\frac{V_{jl}^{D} V_{(t+3)l}^{*D}}{m_{\tilde d_l}^2} \right)
(\bar \nu_i C P_L l_r) (\bar d_k P_L u_m) + \nonumber \\
&&\left. \lambda'_{ijk} \lambda^{'*}_{rst} 
\left(\frac{V_{(k+3)l}^{D} V_{(t+3)l}^{*D}}{m_{\tilde d_l}^2} \right)
(\bar l_r \gamma^\mu P_L \nu_i) (\bar u_m \gamma_\mu P_L d_j) 
\right]
\eea
where $C$ is the charge conjugation operator, $P_L$ is the left projection
operator, $V^D$ is down-squark rotation matrix and $V^L$ is the slepton rotation
matrix. The second coefficient disappears for each of the processes we are 
considering due to the antisymmetric nature of $\lambda_{ijk}$. Furthermore
in the limit when the left-right mixing elements in the slepton and 
down-squark mass matrices are small, the diagonalization matrices $V^D$ and $V^L$ are the identity. Hence the 4-fermi interaction
terms reduce to
\bea
\mathcal{L}_{4f} &\subset& -  V_{sm}^{\rm CKM} \left[
\left(\frac{\lambda_{3j3} \lambda^{'*}_{js3}}{m_{\tilde l_j}^2}
\right) (\bar \tau P_L \nu_\tau) (\bar u_m P_L b) +   
\left(\frac{\lambda'_{33k} \lambda^{'*}_{3sk}}{m_{\tilde d_k}^2} \right)
(\bar \tau \gamma^\mu P_L \nu_\tau) (\bar u_m \gamma_\mu P_L b) 
\right] + {\rm h.c.} 
\eea
where 
we have only kept the leading terms in $\lambda'$. We note that the first term has
exactly the same form as the operator induced by the charged Higgs exchange in 
the MSSM and the second term has an identical structure to the Standard Model. 
A combined analysis of the contributions from both R-parity violating terms and 
the charged Higgs terms have previously been studied in the context of $B \to \tau \nu$ 
in Ref.~\cite{Aida:2010xi,Bose:2011tb}. However we shall assume the charged Higgs 
mass $m_{H^\pm}$ is large and 
therefore the charged Higgs contribution can be neglected. Notice that the 
slepton
contribution is suppressed compared to the squark contribution due to the 
$\lambda'_{3ik} \gg \lambda'_{jik}$ because $j \neq 3$ for non-vanishing 
$\lambda$. Furthermore as the slepton is 
purely left-handed, if its contribution to $B \to \tau \nu$
interferes constructively with the Standard Model, then its contribution to 
$R(D)$ must necessarily be destructive. Therefore it is difficult for slepton 
contribution to simultaneously explain the increase in $B \to \tau \nu$, $R(D)$
and $R(D^*)$~\cite{Crivellin:2012ye}. However, the second term can enhance all the three
B decay modes 
\bea
L_{\rm EFF} =- \frac{4 G_f}{\sqrt{2}} \sum_{m=1,2} V_{3m}^{\rm CKM} \left[1 + \Delta_m \right] (\bar u_{m} \gamma^\mu P_L b) (\bar \tau \gamma^\mu P_L \nu_\tau)
+ {\rm h.c.}
\eea
where 
\bea
\Delta_m &=& \frac{\sqrt{2} }{4G_f} \frac{\lambda'_{33k}}{2m_{\tilde d_k}^2} \sum_{s=1}^3 \lambda^{'*}_{3sk} \left(\frac{V_{sm}^{\rm CKM}}{V_{3m}^{\rm CKM}}\right) \\
&=&  \frac{\sqrt{2} }{4G_f} \frac{|\lambda'_{333}|^2}{2m_{\tilde d_k}^2} \equiv \Delta. \label{eq:deltadef}
\eea
where the last line follows from our choice of the RPV couplings discussed in
Sec.~\ref{sec:setup}.
Although in our model $\Delta_2 = \Delta_1$, we note that a more
general phenomenology will allow them to be different depending on the magnitudes and phases of 
$\lambda'_{32k}$ and $\lambda'_{31k}$ compared to $V^{\rm CKM}$. This can lead to different
enhancements in $B \to D(D^*)\tau \nu$ compared $B \to \tau \nu$. A discussion
of some of the issues arising from such splittings can be found in
Ref.~\cite{Aida:2010xi,Bose:2011tb}. There are weak bounds on
$|\lambda'_{32k}| \lsim 0.52 \, m_{\tilde d_k}/100(\rm GeV)$ and $|\lambda'_{321}| \lsim 0.12 \, m_{\tilde d_k}/100(\rm GeV)$~\cite{Barbier:2004ez}.  However we do not pursue this possibility. 

\begin{figure}
\centering
\includegraphics[width=0.4\textwidth]{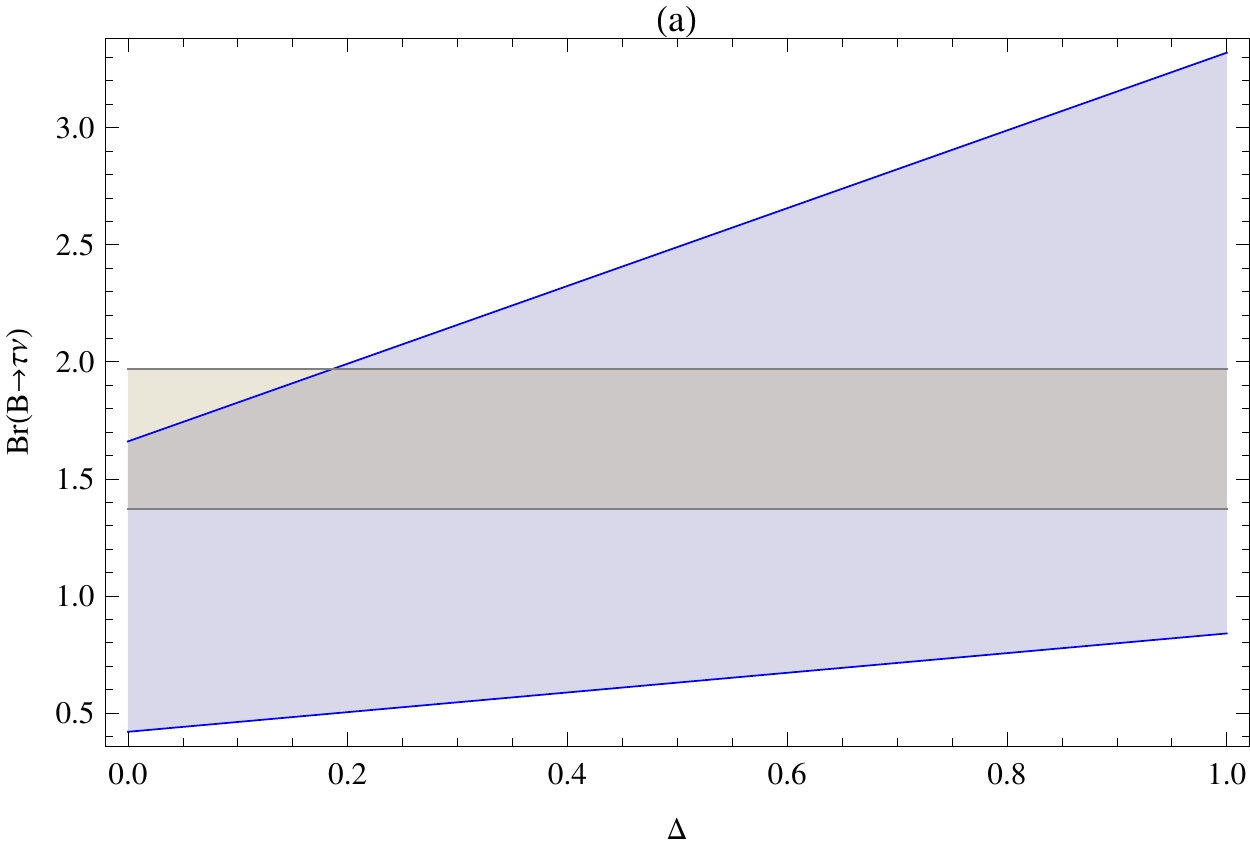}
\includegraphics[width=0.4\textwidth]{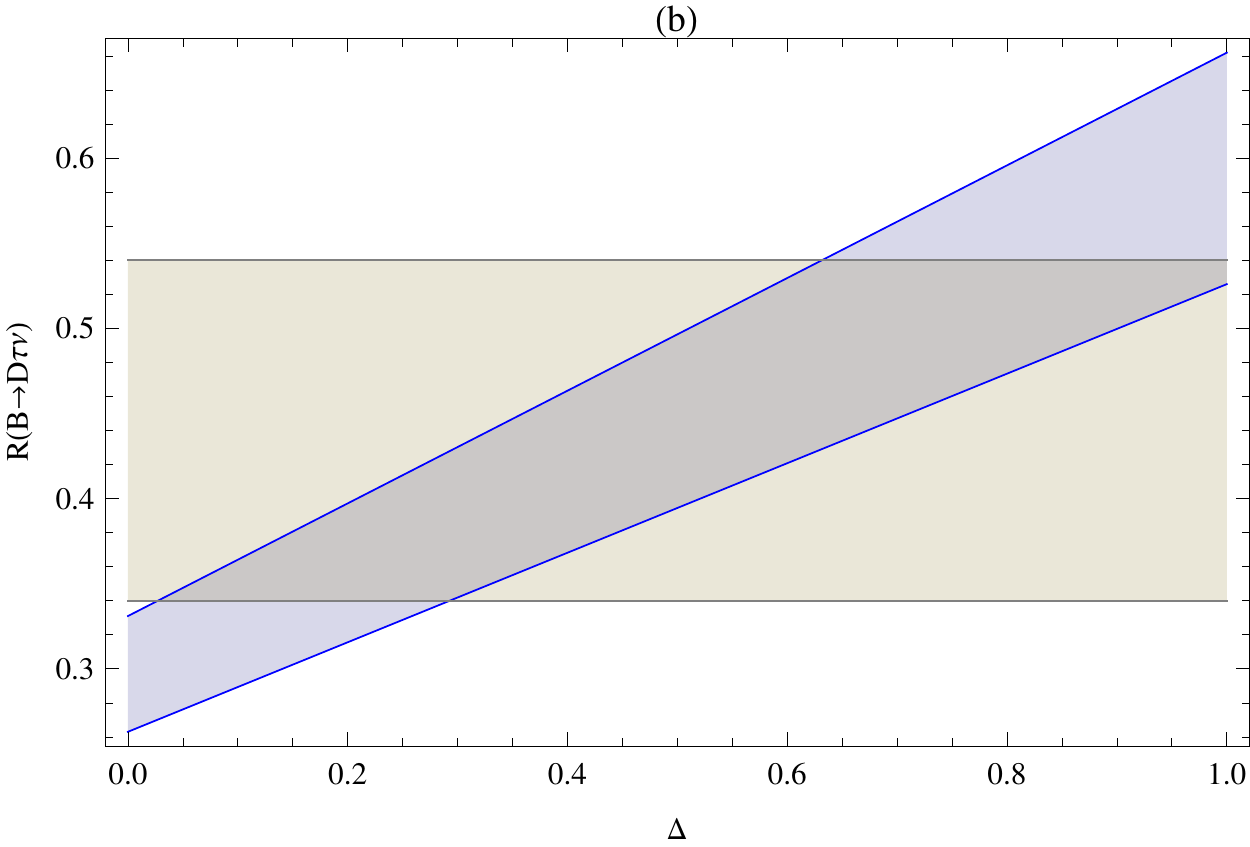}\\
\includegraphics[width=0.4\textwidth]{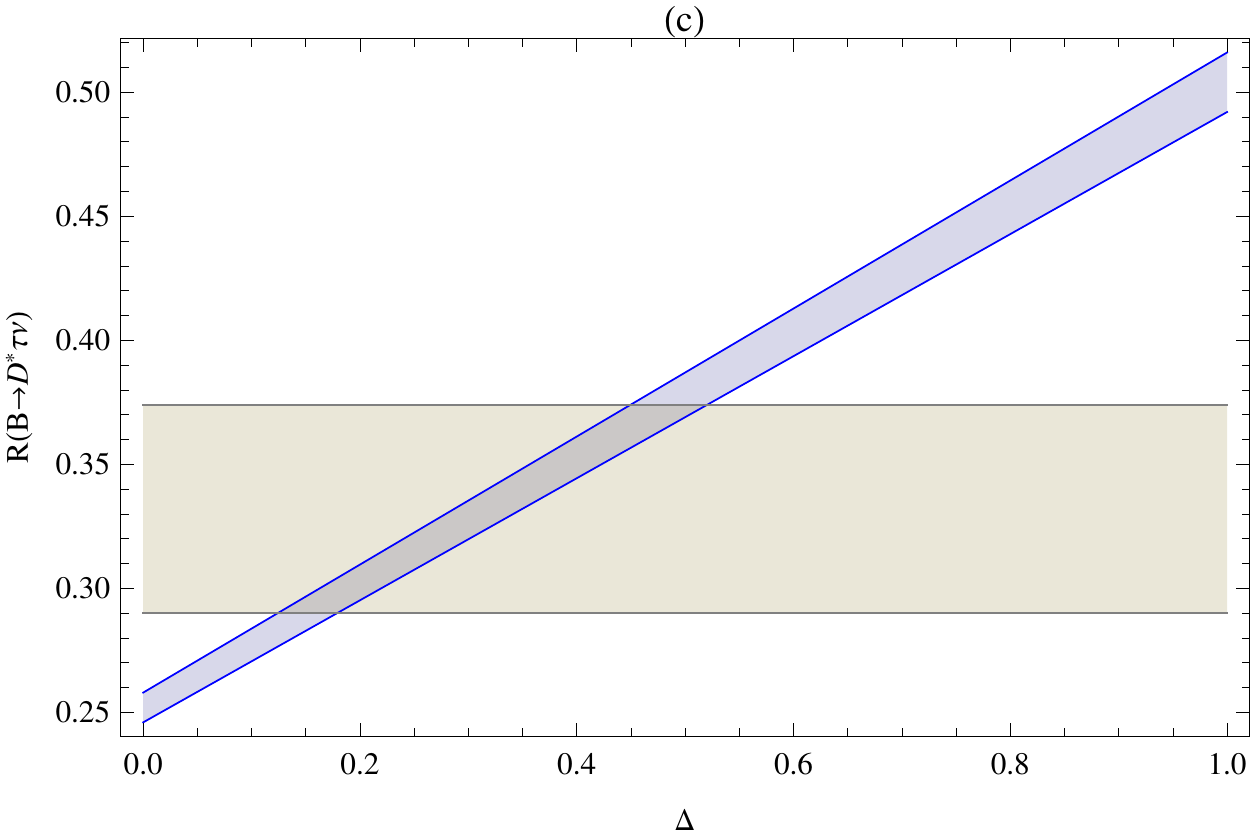}
\label{fig:deltadep}
\caption{Dependence on $\Delta$ of the three observables: (a) 
$\mathcal{BR}(B^+ \to \tau \nu)$, (b) $R(D)$ and (c) $R(D^*)$. The horizontal gray band
corresponds to the $2\sigma$ experimentally allowed values, while the blue band 
corresponds predicted values in RPV supermmetric models assuming the $2\sigma$
uncertainties in the Standard Model predictions.}
\end{figure}

In Fig.~\ref{fig:deltadep} we show dependence of $\mathcal{R}(B \to \tau \nu)$,
$R(D)$ and $R(D^*)$ on $\Delta$, respectively. The horizontal grey bands correspond to the
$2\sigma$ experimentally allowed regions, while the blue bands correspond to 
the predicted values for each of these observables assuming the $2\sigma$
uncertainty in the SM calculated value. In particular we see that the $R(D^*)$
observable puts the strongest constraints on $\Delta$ and leads to the 
constraint that 
\bea
0.12 \lsim \Delta \lsim 0.52.
\label{eq:delta_lim}
\eea
Our proposal leaves the distribution of the $\tau$-lepton and the $D^*$ 
polarization the same as in the Standard Model. Observing 
polarization that differ from the Standard Model would be a way of 
distinguishing our model from other suggestions.

\subsection{Observing this scenario at LHC}

\begin{figure}
\centering
\includegraphics[width=0.4\textwidth]{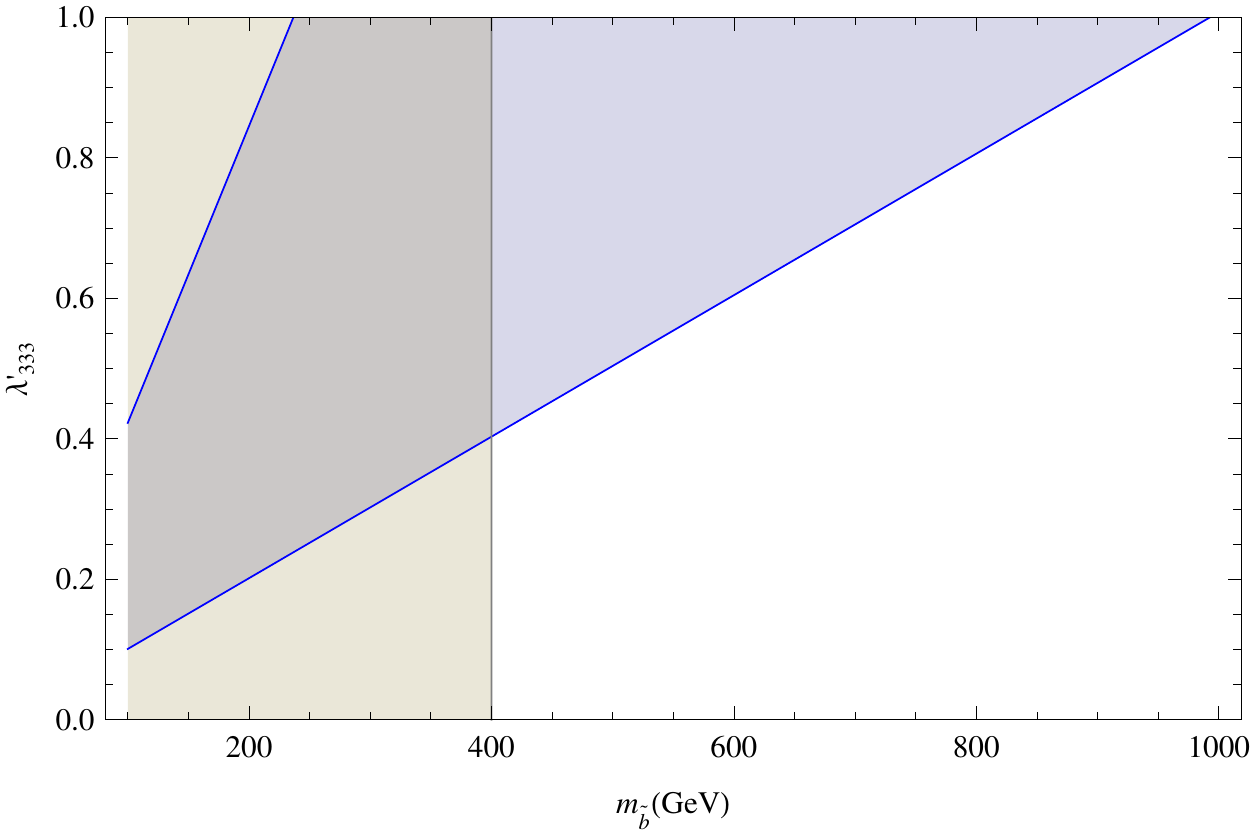}
\caption{The blue (dark gray) region of the $\lambda'_{333}$ vs. $m_{\tilde b_R}$ 
parameter space can explain the observed enhancement in $B$ semi-leptonic 
and leptonic decays. The vertical grey region is excluded by the ATLAS search
for sbottoms.}
\label{fig:lambdavsm}
\end{figure}

Assuming that the enhancement of the observables above their Standard Model
values is purely due to RPV supersymmetry involving the third generation,
the dominant production mechanisms for the colored SUSY particles are 
unmodified. However the decays of the 
stops, sbottoms and staus can be significantly modified. In particular, as a
large $\lambda'_{333}$ is needed to explain the enhanced decays of the $B$ 
mesons the decays $\tilde t \to b l^+$ and $\tilde b \to b \nu$ would compete 
with their standard SUSY decay channels.

For stops, the $gg \to \tilde{t} \tilde{t}^* \to b \bar{b} \tau^+ \tau^-$ 
channel is similar to the Standard Model $gg \to t\bar t \to b \bar{b}\tau^+ \tau^- 
\nu_\tau \bar \nu_\tau$ channel. Hence for stop masses $m_{\tilde t} \lsim 200
$~GeV this search mode could be quite challenging. The mass of the lightest
stop is not crucial in explaining the three B-physics observables we consider.
Therefore the CMS limit on the stop mass $m_{\tilde t_1} \gsim 450
$~GeV~\cite{CMSstop} in 
such a scenario for $\lambda_{333} \sim \mathcal{O}(1)$ does not put a strong
constraint on our model.

For sbottoms, the $gg \to \tilde b \tilde b^* \to b \bar{b} \nu_\tau \bar 
\nu_\tau$ is similar to the standard SUSY search $\tilde b \to b 
\chi_0^1$ with $m_{\chi_0^1} = 0$. Hence the ATLAS limits~\cite{Aad:2011cw}
$m_{\tilde b} \gsim 400$~GeV are relevant to this study. This limit assumes that
the $\tilde b $ decay predominantly to $b \nu_\tau$ which is not the case for 
heavier sbottom masses where the $\tilde b \to \bar t \tau$ will also be open. 
Therefore the ATLAS limit on the sbottom mass may be slightly weaker. In 
Fig.~\ref{fig:deltadep} we present a combination of the conservative constraint
from ATLAS on the sbottom mass and from Eq.~(\ref{eq:delta_lim}) for the case 
where only $\lambda'_{333} \neq 0$.
In Fig.~\ref{fig:deltadep} the vertical grey region is excluded by the ATLAS
search, while the blue (dark grey) area is the allowed region of parameter 
space in 
the $\lambda'_{333} \-- m_{\tilde b_R}$ plane. In particular a large RPV couplings
is needed in order to explain the observed enhanced decay rates of $B^+ \to
\tau \nu$, $B \to D \tau \nu $ and $B \to D^{(*)} \tau \nu$. Furthermore,
future searches in $b \bar{b}+$MET channel at LHC will be able to probe
a significant portion of the parameter space of this scenario. The 
additional decay mode $\tilde b \to b l^- \tau^+ \nu_\tau$ can also be an 
interest mode for discovery of this scenario, especially considering the 
multilepton searches at CMS~\cite{Chatrchyan:2012mea}.

\section{Conclusions}
\label{sec:concl}

We have proposed a model that gives a universal explanation of the 
experimentally observed enhancements in in the 
leptonic and semi-leptonic decays of the B meson into the third generation of 
leptons. We invoke R-parity violating operators in the the MSSM.
Only the operator $\lambda'_{ijk} L_i Q_j D_k $ is necessary, and the effective 
interaction resulting from d-squark exchange has the same form as the SM 
current-current interaction. We estimate the strength of the new interaction 
using a fit to the data, and derive bounds on the d-squark mass assuming 
perturbative unitarity holds for the Yukawa couplings. We then discuss briefly 
the production and signatures of d-squark in this model. We conclude that the 
model is testable at the LHC.

\textbf{Acknowledgments:}
  N.G.D.\ and A.M.\ are supported by the U.S.\ Department of Energy
  under Contract No. DE-FG02-96ER40969

\end{document}